\documentclass{aa}  

\usepackage{amssymb}
\usepackage[usenames, dvipsnames]{color}
\usepackage{graphicx}
\usepackage{natbib}
\usepackage{nicefrac}
\usepackage{sidecap}
\usepackage{txfonts}
\usepackage{units}
\usepackage{url}

\graphicspath{{//home/meetu/article6/Astroph/}}
\usepackage{hyperref}
\hypersetup{
    final=true,
    pageanchor=true,
    colorlinks=true,
    breaklinks=true,
    linkcolor=blue,
    citecolor=blue,
    urlcolor=blue,
    pdfpagemode=UseNone,
    pdftitle={Evaluating local correlation tracking using CO5BOLD simulations
             of solar granulation},
    pdfauthor={},
    pdfsubject={Solar Physics},
    pdfkeywords={Sun: granulation, convection, hydrodynamics, 
                 methods: numerical,
                 techniques: image processing}}

\begin{document} 


\title{Evaluating local correlation tracking\\
    using CO5BOLD simulations of solar granulation}

\subtitle{}

\author{M.\ Verma, M.\ Steffen, \and C.\ Denker}

\institute{Leibniz-Institut f\"ur Astrophysik Potsdam,
    An der Sternwarte 16,
    14482 Potsdam,
    Germany\\
    \email{mverma@aip.de, msteffen@aip.de, cdenker@aip.de}}

\date{Received April 2, 2013 / Accepted May 21, 2013}

 
\abstract
{Flows on the solar surface are intimately linked to solar activity, and local
correlation tracking (LCT) is one of the standard techniques for capturing the
dynamics of these processes by cross-correlating solar images. However, the
link between contrast variations in successive images to the underlying plasma
motions has to be quantitatively confirmed.}
{Radiation hydrodynamics simulations of solar granulation (e.g.,
CO$^\mathrm{5}$BOLD) provide access to both the wavelength-integrated, emergent
continuum intensity and the three-dimensional velocity field at various heights
in the solar atmosphere. Thus, applying LCT to continuum images yields
horizontal proper motions, which are then compared to the velocity field of the
simulated (non-magnetic) granulation. In this study, we evaluate the performance
of an LCT algorithm previously developed for bulk-processing \textit{Hinode}
G-band images, establish it as a quantitative tool for measuring horizontal
proper motions, and clearly work out the limitations of LCT or similar
techniques designed to track optical flows.}
{Horizontal flow maps and frequency distributions of the flow speed were
computed for a variety of LCT input parameters including the spatial resolution,
the width of the sampling window, the time cadence of successive images, and the
averaging time used to determine persistent flow properties. Smoothed velocity
fields from the hydrodynamics simulation at three atmospheric layers ($\log\tau
= -1$, 0, and $+1$) served as a  point of reference for the LCT results.}
{LCT recovers many of the granulation properties, e.g., the shape of the flow
speed distributions, the relationship between mean flow speed and averaging
time, and  also -- with significant smoothing of the simulated velocity field --
morphological features of the flow and divergence maps. However, the horizontal
proper motions are grossly underestimated by as much as a factor of three. The
LCT flows match best the flows deeper in the atmosphere at $\log \tau = +1$.}
{Despite the limitations of optical flow techniques, they are a valuable
tool in describing horizontal proper motions on the Sun, as long as the results
are not taken at face value but with a proper understanding of the input
parameter space and the limitations inherent to the algorithm.}


\maketitle


\section{Introduction}


%

The interaction of flow and magnetic fields on the solar surface is 
linked to the ever-changing appearance of solar activity. Various techniques
have been developed to measure photospheric proper motions. \citet{November1988}
introduced the local correlation tracking (LCT) technique to the solar physics
community, which had previously been developed by \citet{Leese1971} to track
cloud motion from geosynchronous satellite data. Since then many variants of the
LCT algorithm have been used to quantify horizontal flow properties in the solar
photosphere and chromosphere. 
\citet{Verma2011} adapted the LCT algorithm to G-band images captured by the
Solar Optical Telescope \citep[SOT,][]{Tsuneta2008} on board the Japanese
\textit{Hinode} mission \citep{Kosugi2007} with the aim for establishing a
standard method for bulk-processing 
time-series data. \citet{Beauregard2012} modified the same algorithm for
continuum images of the Helioseismic and Magnetic Imager
\citep[HMI,][]{Scherrer2012} on board the Solar Dynamics Observatory
\citep[SDO,][]{Pesnell2012}.

Besides images, magnetograms are often used to determine horizontal proper
motions. \citet{Welsch2007} tested and compared various techniques:
minimum-energy fitting \citep[MEF,][]{Longcope2004},  the differential affine
velocity estimator \citep[DAVE,][]{Schuck2006},  Fourier local correlation
tracking \citep[FLCT,][]{Welsch2004},  the induction method
\citep[IM,][]{Kusano2002}, and  induction local correlation tracking
\citep[ILCT][]{Welsch2004}. These algorithms were applied to simulated
magnetograms using the anelastic MHD code by \citet{Lantz1999}. However, the
results were not conclusive, because all methods showed considerable errors in
estimating velocities. In general, MEF, FLCT, DAVE, IM, and ILCT lead to similar
results with a slightly better performance of  DAVE  in estimating direction and
magnitude of the velocities, but the magnetic flux energy and helicity were
recovered better using MEF.

The study by \citet{Welsch2007} was extended by \citet{Chae2008}, who tested
three optical flow methods, i.e., LCT, DAVE, and the nonlinear affine velocity
estimator \citep[NAVE,][]{Schuck2005}, using simulated, synthetic, and
\textit{Hinode} magnetograms. NAVE performed well in detecting subpixel,
superpixel, and nonuniform motions, whereas LCT had problems in sensing
non-uniform motions, and DAVE displayed deficiencies in estimating superpixel
motions.  However, LCT is the fastest and NAVE the slowest algorithm.
\citet{Chae2008} proposes to  select smaller sampling windows to get more
detailed velocity maps and to lessen the computational demands.

\citet{Rieutord2001} applied LCT and feature tracking to photospheric images
derived from numerical simulations of compressible convection and concluded
that both methods fail to represent the velocity field on spatial scales smaller
than 2500~km and temporal scales shorter than 30~min. In addition, simulated
continuum images \citep{Vogler2005b} were used by \citet{Matloch2010} to
characterize properties of mesogranulation. However, to the best of our
knowledge no systematic study has been carried out based on simulated continuum
images to evaluate the reliability, accuracy, and parameter dependence of LCT.
In this study, we present the results of rigorously testing the LCT algorithm of
\citet{Verma2011} using CO$^\mathrm{5}$BOLD simulations of solar convection
\citep{Freytag2012}. Ultimately, we want to answer the question: How much of the
underlying physics can be captured using optical flow methods?


\section{CO$^\mathrm{5}$BOLD simulation of solar granulation}

Radiation hydrodynamic simulations of solar and stellar surface convection have
become increasingly realistic producing many aspects of
observations. The CO$^\mathrm{5}$BOLD code \citep{Freytag2012} offers a unique
opportunity to evaluate a previously developed LCT algorithm \citep{Verma2011}
and to explore the parameter space for tracking solar fine-structure, in
particular for scrutinizing the multi-scale (time and space) nature of solar
surface convection. CO$^\mathrm{5}$BOLD simulations have been computed for a
variety of solar models. The present CO$^\mathrm{5}$BOLD simulations are
non-magnetic, i.e., pure radiation hydrodynamics. Here, the grid dimensions of
the simulation are $400 \times 400 \times 165$ and the horizontal cell size is
28~km $\times$ 28~km with the vertical grid spacing increasing with depth from
12~km in the photosphere to 28~km in the lower part of the model, resulting in a
box with a size of $11.2 \times 11.2 \times 3.1$~Mm$^{3}$. Even larger
simulation boxes are needed to study convective signatures on larger spatial
scales. \citet{Matloch2010}, for example, extracted characteristic properties of
mesogranulation (e.g., size and lifetime) from three-dimensional hydrodynamical
simulations \citep[MURaM code,][]{Vogler2005b}.

\begin{figure}
\includegraphics[width=\columnwidth]{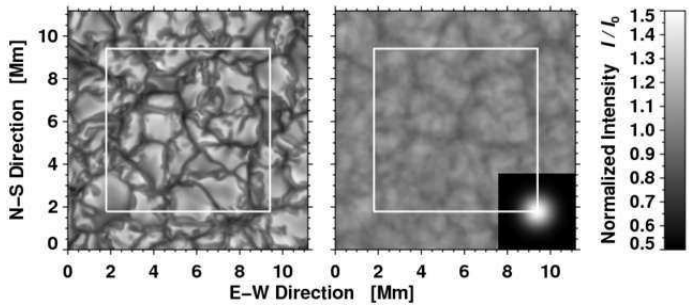}
\caption{First (\textit{left}) and one-hour averaged (\textit{right}) images
with 400 $\times$ 400 pixels and an image scale of 28~km pixel$^{-1}$ taken from
a CO$^\mathrm{5}$BOLD time-series of solar granulation. Because of the Gaussian
kernels (\textit{insert in the lower right corner}) of 128 $\times$ 128 pixels
used as sampling windows, the original image size was reduced by 128~pixels. The
white squares mark the region with a size of 273 $\times$ 273 pixels that
remains after the LCT computation.}
\label{FIG01}
\end{figure}

In the first part of this study, we analyze a time-series of
snapshots of the bolometric emergent continuum intensity based on a
high-resolution CO$^\mathrm{5}$BOLD simulation of granulation. The 140-minute
time series comprises 832 images with a size of $400 \times 400$ pixels at a
time cadence of $\Delta t=10$~s. The image scale is 28~km pixel$^{-1}$.  The
first image of the time series is shown in the left panel of Fig.~\ref{FIG01}.
The intensity contrast of image is about 15.9\%. Taking the average of 360
images (right panel of Fig.~\ref{FIG01}) reduces the contrast to about 5.3\% but
the appearance of the intensity pattern still remains granular without any
indication of long-lived features. In general, the modulation transfer function
(MTF) of the telescope has to be considered when comparing simulated with
observed data \citep{Wedemeyer2009}. However, we did not convolve the images
with the point spread function to include the instrumental image degradation,
because the fine-structure contents of an image are given by
the Fourier phases and not the Fourier amplitudes. Modifying the Fourier
amplitudes in granulation images or changing the shape and width of the LCT
sampling window are intertwined and could be in principle disentangled but at
the cost of complexity. In the second part of the data analysis, we compare the
LCT results with the underlying velocity structure of the CO$^\mathrm{5}$BOLD
simulations. The time cadence of the velocity data is, however, reduced to
$\Delta t=20$~s. 
For an impartial comparison, all LCT maps were computed first
before confronting them with the intrinsic velocity field of the
CO$^\mathrm{5}$BOLD simulations.


\section{Local correlation tracking}

Flow maps were computed using the LCT algorithm described in \citet{Verma2011}
but initial image alignment, subsonic filtering, and geometric correction were
excluded. An important difference between the \textit{Hinode} and
CO$^\mathrm{5}$BOLD data is the image scale of 80 and 28~km pixel$^{-1}$,
respectively. Sampling windows and filter kernels of $32 \times 32$ pixels were
the choice for processing \textit{Hinode} G-band images because the computation
time should be kept to a minimum for bulk-processing of time-series data.
However, for this study we opted for sizes of $128 \times 128$ pixels to adapt
to the higher spatial resolution of the simulated granulation images and to
extend our parameter study to include broader sampling windows. The drawback is
of course a much increased computation time. In principle, smaller sampling
windows could have been chosen for smaller FWHM but for consistency, we decided
not to change window or kernel sizes. In a direct comparison with the previous
work of \citet{Verma2011}, these small differences are of no concern. 

A $128 \times 128$-pixel Gaussian kernel with an FWHM = 1200~km is
shown as an illustration in the lower-right corner of the right panel in
Fig.~\ref{FIG01}. If not otherwise noted, this kernel is used as a high-pass
filter, even though this is not strictly necessary in the absence of strong
intensity gradients (e.g., umbra-penumbra or granulation-penumbra boundaries)
and gentler slopes introduced by the limb darkening. Because of such
sampling windows and kernels the size of the computed LCT maps was reduced to
$273 \times 273$ pixels from the $400 \times 400$ pixels of the original images,
which is indicated in Fig.~\ref{FIG01} by white square boxes.


\section{Results}

\subsection{Persistent flows and the duration of time averages}

\begin{figure}
\includegraphics[width=\columnwidth]{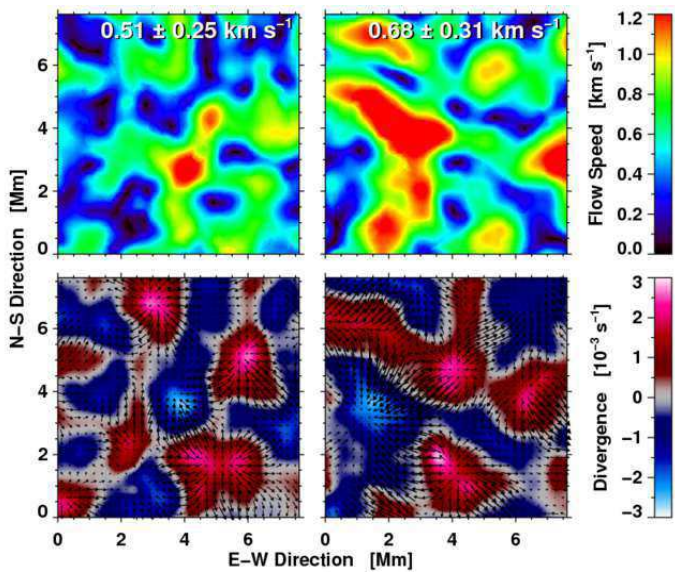}
\caption{Average speed (\textit{top}) and divergence (\textit{bottom}) maps for
the first (\textit{left}) and last (\textit{right}) hour of the time series
computed using a time cadence of $\Delta t = 60$~s and a Gaussian sampling
window with an $\mathrm{FWHM} = 1200$~km. The direction and magnitude of the
horizontal flows are plotted over the divergence maps as arrows, for which a
velocity of 0.5~km~s$^{-1}$ corresponds to exactly the grid spacing. The values
displayed at the top are the mean speed $\bar v$ and its standard deviation
$\sigma_v$ for the one-hour averaged flow maps.}
\label{FIG02}
\end{figure}

\begin{figure*}[t]
\includegraphics[width=\textwidth]{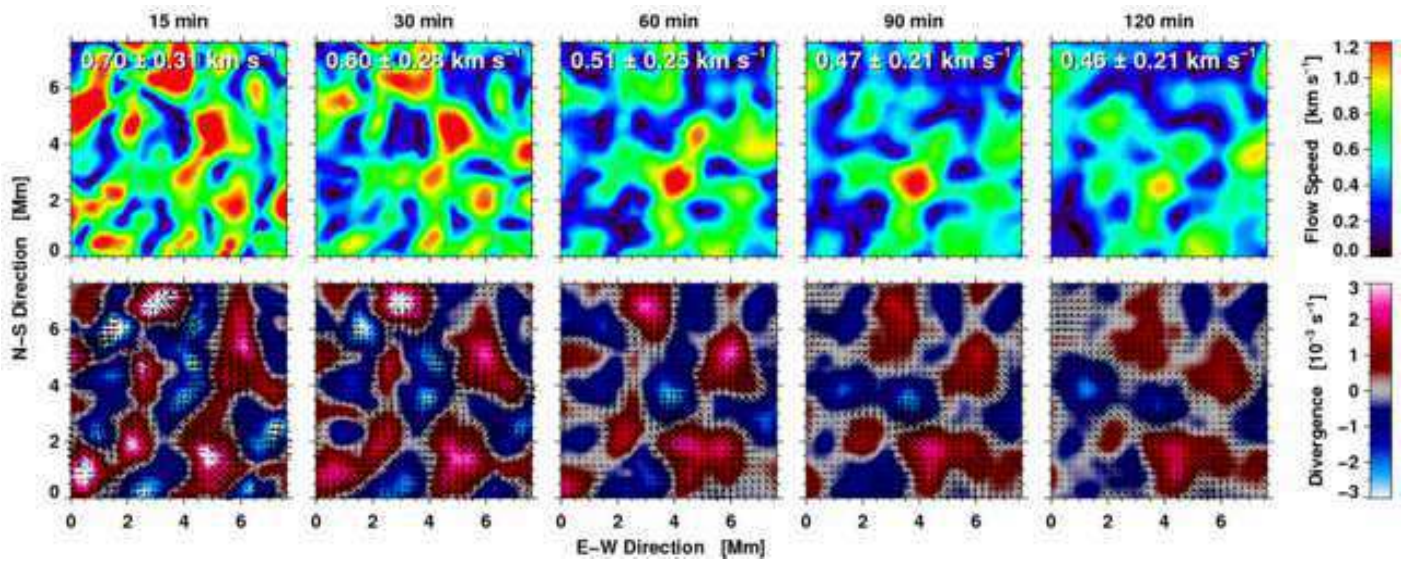}
\caption{Speed (\textit{top}) and divergence (\textit{bottom}) maps averaged
over time intervals of $\Delta T =15$--120~min. The
LCT maps were computed using a time cadence of $\Delta t = 60$~s and a Gaussian
sampling window with an $\mathrm{FWHM} = 1200$~km.}
\label{FIG03}
\end{figure*}

One application of LCT techniques is to uncover persistent horizontal proper
motions. The imprint of solar convection results in a photospheric flow field
with different spatial scales: granulation 1--2~Mm, mesogranulation 3--10~Mm,
and supergranulation $\sim$30~Mm \citep{Muller1992}. The LCT method recovers the
corresponding flow features from time-series of photospheric images. This raises
the question, if similar signatures are present in the simulated time-series of
granulation.

In Fig.~\ref{FIG02}, we computed flow (top row) and divergence (bottom row) maps
for the first (left column) and last (right column) hour of the time series.  We
chose a time cadence of $\Delta t = 60$~s and a Gaussian sampling window with an
$\mathrm{FWHM} = 1200$~km and then averaged the individual flow maps over
$\Delta T = 1$~h to facilitate easier comparison with the study of
\citet{Verma2011}. This choice of input parameters produced the most reliable
results when previously applied to \textit{Hinode} G-band images. The mean
speeds $\bar{v}$ and their standard deviations $\sigma_v$ are $0.51 \pm
0.20$~km~s$^{-1}$ and $0.68 \pm 0.31$~km~s$^{-1}$ for the first and last hour,
respectively. The mean speed in the later case is higher by 33\% because of a
large structure with high velocities in the top-left quarter of the flow map.
The divergence of a flow field with velocity components $(v_x, v_y)$ is computed
as $\nabla \cdot \vec{v} = \nicefrac{\partial v_x}{\partial x} +
\nicefrac{\partial v_y}{\partial y}$. On average the divergence values are two
times higher than the ones for \textit{Hinode} G-band images discussed in
\citet{Verma2011}. The superimposed flow vectors clearly indicate the locations
of the diverging and converging flows.

Neither the two flow nor the two divergence maps bear any resemblance to another
suggesting that after about one hour persistent flow features are no longer
present in the CO$^\mathrm{5}$BOLD time series. The absence of meso- or
supergranular features is, however, expected considering both the horizontal
dimensions and the depth of the simulation box. Interestingly, strong flow
features will still be present even after averaging for one hour, which points
to the necessity of longer time series for assessing convective flow properties
on larger spatial scales.

The time dependence of the average flow field is depicted in Fig.~\ref{FIG03}
along with the corresponding divergence maps. These maps were computed with the
same input parameters as in Fig.~\ref{FIG02} but in this case with different
time intervals  $\Delta T=15$--120~min over which the individual flow maps were
averaged. Proper motions of single granules are well captured when flow maps are
averaged over $\Delta T=15$ and 30~min, which is evident from the speed maps
containing mostly features of typical granular size. However, contributions to
the flow field from separate granules are averaged out over longer time
intervals $\Delta T$, but some long-lived features might survive the averaging
process. Most notably, a high-velocity feature with negative divergence
(converging flows) remains in the central part of the maps. It first appears at
15~min as an appendage to a stronger flow kernel, then separates from the kernel
(30--60~min), before becoming the only strong flow kernel at 90~min, and fading
away after about 120~min. This behavior resembles flow patterns encountered at
the vertices of supergranular cells. However, in real observations the lifetime
of such a flow kernel with converging horizontal motions is much longer. 
Despite the tendency of lower divergence values for increasing $\Delta T$, the
divergence values are still significantly higher than the values given by
\citet{Verma2011}.

The more than 30\% decrease of the average flow speed from 0.70~km~s$^{-1}$ in
the 15-minute map to 0.46~km~s$^{-1}$ in the 120-minute map is of the same order
as the 25\% difference found for the two one-hour averaged flow maps in
Fig.~\ref{FIG02}. Furthermore, the extended flow feature in the right panel of 
Fig.~\ref{FIG02} is absent in the 120-minute map of  Fig.~\ref{FIG03} pointing
to the importance of the vectorial summation (constructive and destructive) of
the individual flow maps. Taken together, both facts serve as a reminder that we
must treat the mean speed values of different sequences with caution when
comparing them. In addition, the large variance of reported speed values in
the literature might also have its origin in the stochastic nature of granulation
and not only in the choice of LCT input parameters. In the following, only data
for the first hour of the time series were analyzed.

\subsection{Convergence properties of the mean flow speeds}              

\textit{Hinode} is the most prolific data source for high-resolution solar
images, and  G-band images with $2 \times 2$-pixel binning and an image scale of
about 80~km pixel$^{-1}$ are the most common image type.  Hence, the LCT
algorithm of \citet{Verma2011} was developed for bulk-processing of these images
leading to an optimal set of LCT input parameters: time cadence $\Delta
t=60$--90~s, Gaussian sampling window with an $\mathrm{FWHM}=1200$~km,  and
averaging time $\Delta T=1$~h. \citet{Beauregard2012} showed that this algorithm
is easily adaptable to images with a coarser spatial sampling, e.g., continuum
images of the SDO/HMI. In the following, we use CO$^\mathrm{5}$BOLD simulation
data to cross-check the aforementioned input parameters. 

\begin{figure}
\includegraphics[width=\columnwidth]{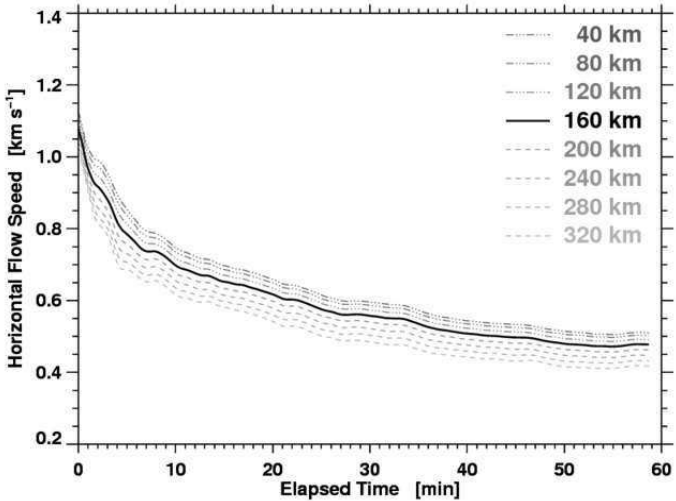}
\caption{Average horizontal flow speed as a function of the elapsed time. The
maps were computed using a Gaussian sampling window with an $\mathrm{FWHM} =
1200$~km and time cadence $\Delta t = 60$~s. Before computing the
cross-correlations, the images were convolved with a Gaussian kernel
$\mathrm{FWHM} = 40$--320~km to degrade the spatial resolution of the images.
The increasing FWHM values are presented as progressively lighter shades of
gray. The dash-dotted and dashed lines represents values above
and below optimal parameter ($\mathrm{FWHM} = 160$~km), respectively.
The black solid line is used to depict result for the $\mathrm{FWHM} = 160$~km,
which is the typical spatial resolution of most  \textit{Hinode} G-band images
\citep[see][]{Verma2011}.}
\label{FIG04}
\end{figure}

\begin{figure}
\includegraphics[width=\columnwidth]{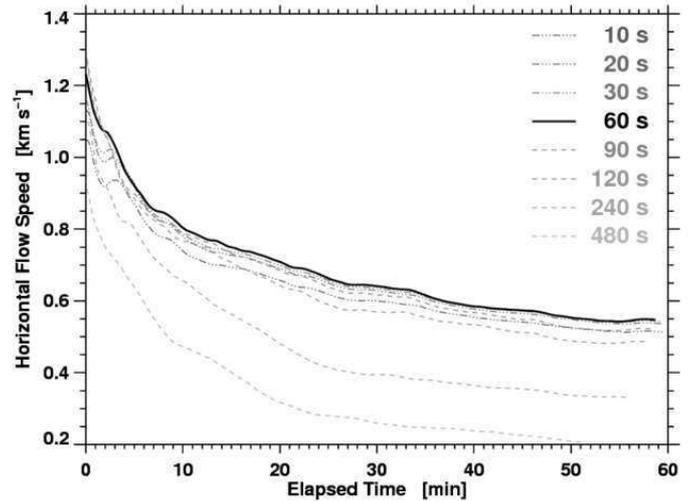}
\caption{Average horizontal flow speed as a function of the elapsed time. The
maps were computed using a Gaussian sampling window
with an $\mathrm{FWHM} = 1200$~km and time cadence $\Delta t = 10$--480~s. The
increasing time cadences $\Delta t $ are presented as progressively lighter
shades of gray. The black solid line is depicting the result for the time
cadence $\Delta t = 60$~s, which has been identified in \citet{Verma2011} as an
optimal choice for tracking horizontal proper motions in \textit{Hinode} G-band
time-series data.}
\label{FIG05}
\end{figure}

\begin{figure}
\includegraphics[width=\columnwidth]{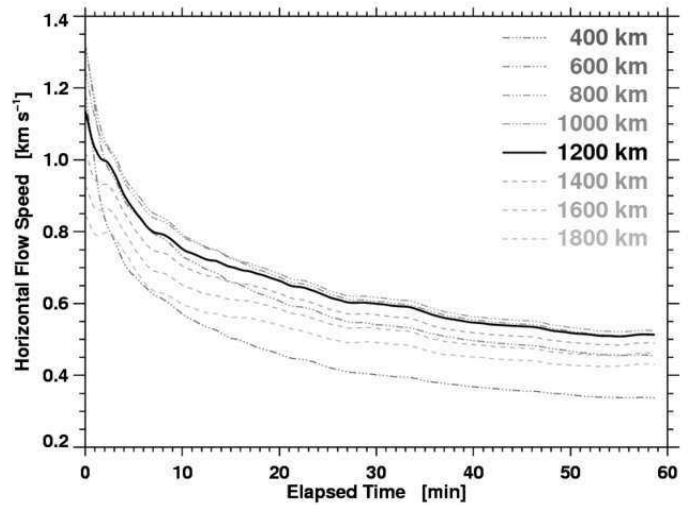}
\caption{Average horizontal flow speed as a function of the elapsed time. The
maps were computed using a time cadence $\Delta t =
60$~s and Gaussian sampling window with various $\mathrm{FWHM} = 400$--1800~km.
The increasing FWHM values are presented as progressively lighter shades of
gray. The black solid line is used to depict the result for a Gaussian sampling
window with an $\mathrm{FWHM} = 1200$~km, which has been identified in
\citet{Verma2011} as an optimal choice for tracking horizontal proper motions in
\textit{Hinode} G-band time-series data.}
\label{FIG06}
\end{figure}

\begin{figure*}
\includegraphics[width=0.96\textwidth]{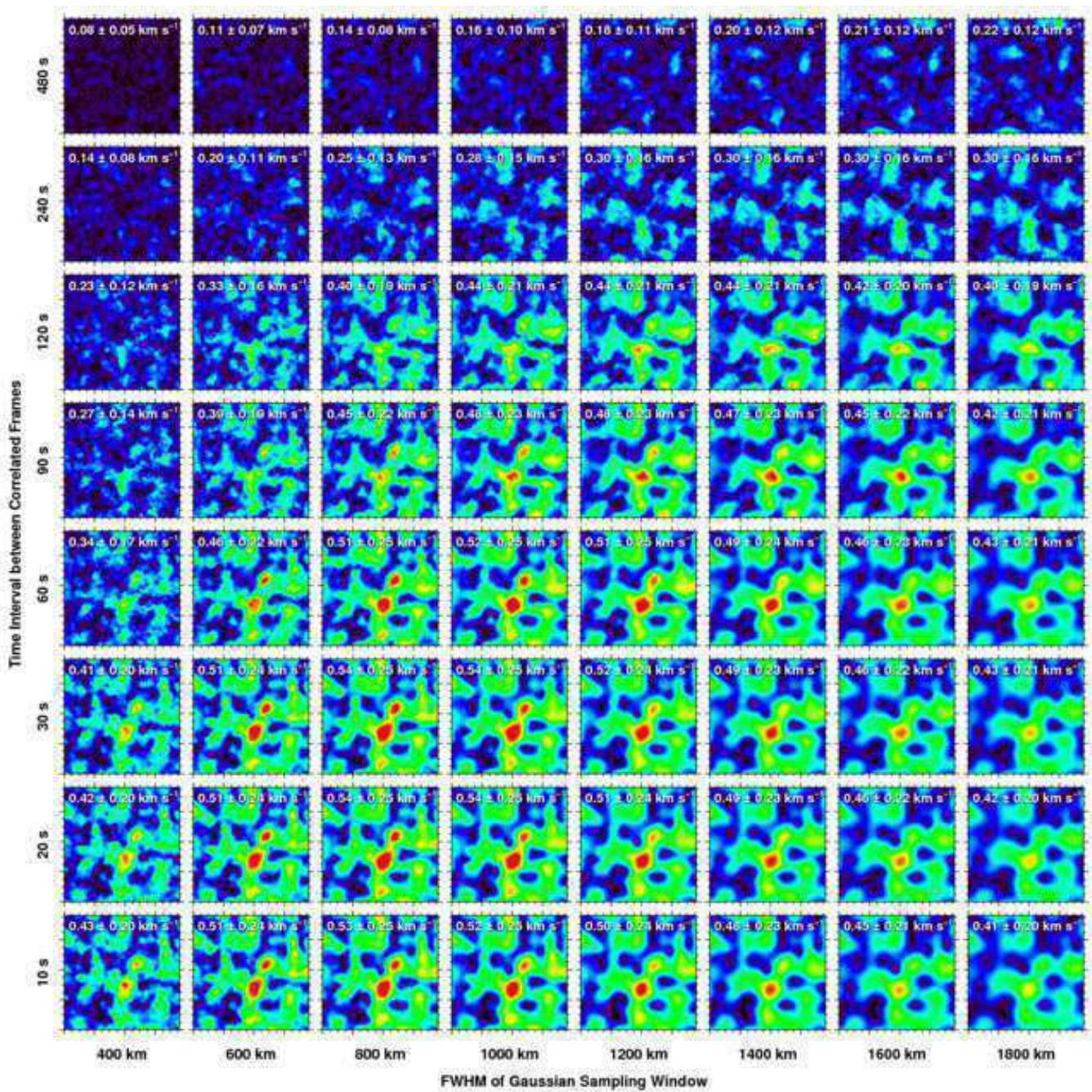}
\caption{Horizontal flow speed maps (one-hour averages) computed with time
cadences $\Delta t = 10$--480~s (\textit{bottom to top}) and Gaussian kernels
with an $\mathrm{FWHM} = 400$--1800~km (\textit{left to right}). Displayed at
the top of the panels are the mean speed $\bar{v}$ and its standard deviation
$\sigma_v$ averaged over the FOV. The speed maps are scaled between
0.0--1.2~km~s$^{-1}$, which corresponds to the scale used in Figs.~\ref{FIG02}
and \ref{FIG03}. Axes are in megameters with the major tickmarks separated by
2~Mm.}
\label{FIG07}
\end{figure*}

First, to study the effects of spatial resolution, we degraded the original
images with an image scale of 28~km pixel$^{-1}$ by convolving them with
Gaussian kernels of $\mathrm{FWHM}=40$--320~km but without correcting the
telescope's MTF \citep[cf.,][]{Wedemeyer2009}. We computed the mean horizontal
flow speed as a function of the elapsed time, as shown in Fig.~\ref{FIG04}. The
elapsed time corresponds to the number of flow maps averaged. These temporal
averages were carried out up to $\Delta T=1$~h. The Gaussian sampling window
with an $\mathrm{FWHM}=1200$~km and time cadence $\Delta t=60$~s were kept
constant. Different shades of gray from dark to light indicate increasingly
larger FWHM. The black curve ($\mathrm{FWHM}=160$~km) corresponds to the spatial
resolution of the \textit{Hinode} G-band images. The diffraction-limited
resolution of the 0.5-meter \textit{Hinode}/SOT according to the Rayleigh
criterion is $1.22 \cdot \lambda / D = 0.22\arcsec$ at $\lambda$430~nm or 160~km
on the solar surface. Therefore, even with $2 \times 2$-binning, \textit{Hinode}
G-band images are still critically sampled (Nyquist theorem). In the following,
we use $v_{28}$ to indicate velocities that were derived from
the full-resolution simulated images, while $v_{80}$ refers to velocities based
on smoothed simulated images, which provide the link to G-band images
\citep[see][]{Verma2011}.

\begin{figure*}
\includegraphics[width=0.96\textwidth]{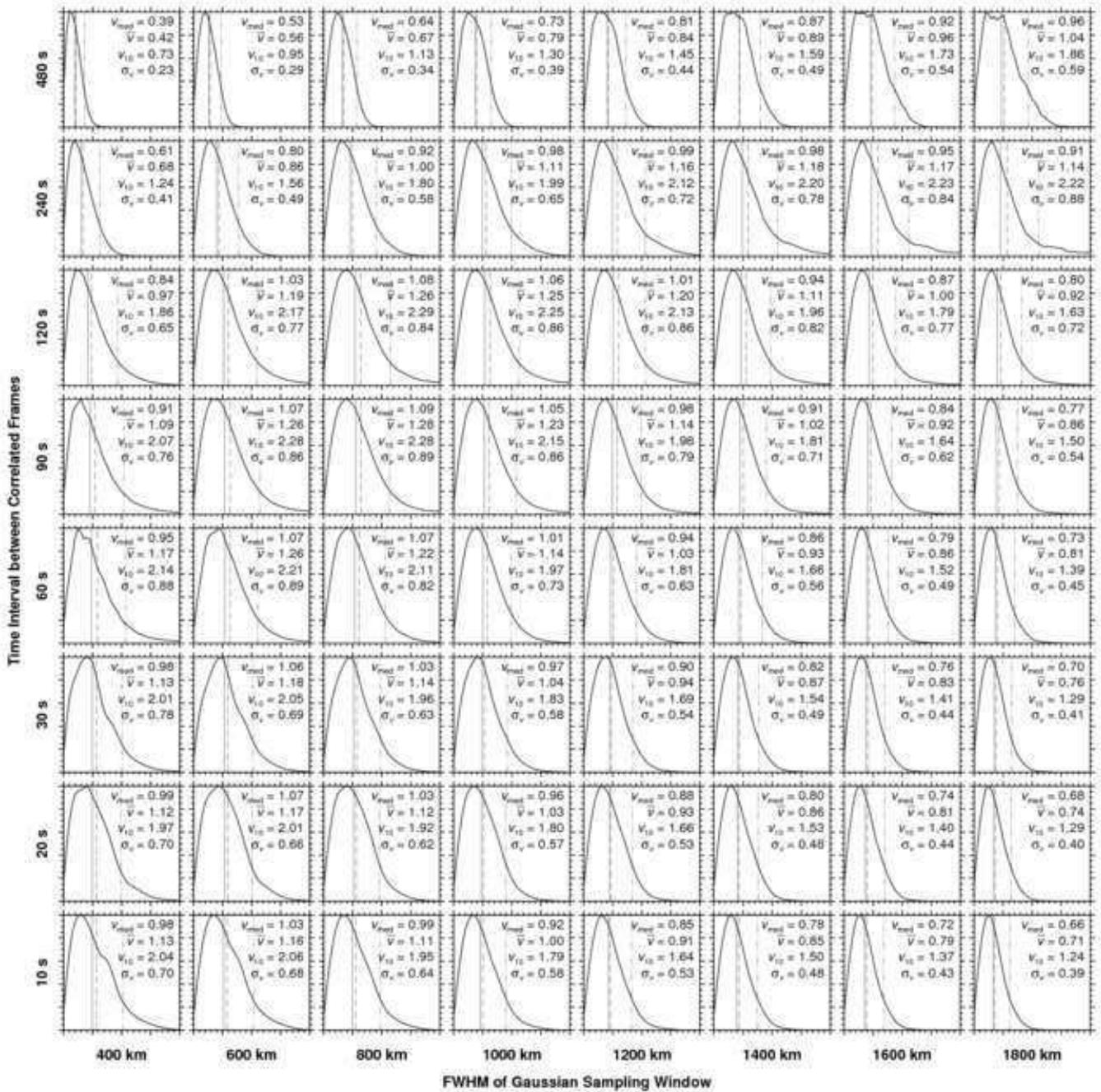}
\caption{Relative frequency distributions of horizontal flow speeds computed
with time cadences $\Delta \mathrm{t} = 10$--480~s (\textit{bottom to top}) and
Gaussian sampling windows with an $\mathrm{FWHM} = 400$--1800~km (\textit{left
to right}). The three vertical lines mark the position of median
$v_\mathrm{med}$ (\textit{solid}), mean $\bar v$ (\textit{long-dashed}), and
$10^\mathrm{th}$ percentile $v_{10}$ (\textit{dash-dotted}) values of speed. The
frequency distributions were normalized such that the modal values correspond to
unity, i.e., major tickmarks are separated by 0.2 on the ordinate. The speed
distributions cover the range 0.0--4.0~km~s$^{-1}$, i.e., major tickmarks are
separated by 1.0~km~s$^{-1}$ on the abscissae.}
\label{FIG08}
\end{figure*}

It takes about 5--10~min before the slope of the curves changes,  and after
15--20~min the curves level out and approach an  asymptotic value. Changing the
slope is directly related to the lifetime of individual granules, and it takes
several lifetimes to reveal any longer-lasting flow features. In addition, all
the curves are stacked on top of each other without crossing any other curve
at any point. The mean flow speed decreases monotonically by about
0.1~km~s$^{-1}$ when reducing the FWHM of the Gaussian filter kernel from 40 to
320~km. Evidently, in images with high spatial resolution, the LCT algorithm
picks up fine-structure such as the corrugated boundaries of granules, 
fragments of exploding granules, and the occasional bright points. These feature
have either intrinsically higher speeds or high contrasts, thus biasing the
cross-correlation algorithm to higher velocities.

Second, we change the time cadences in the range of $\Delta t=10$--480~s while
keeping the $\mathrm{FWHM}=1200$~km of the Gaussian sampling window and the
spatial resolution of 28~km pixel$^{-1}$ constant as depicted in
Fig.~\ref{FIG05}. Shades of gray from dark to light indicate longer time
cadences. The black curve corresponds to the optimal LCT parameter  $\Delta
t=60$~s for G-band images. This curve exhibits the highest mean speed except for
very short averaging times. For shorter time cadences $\Delta t=10$--30~s, the
mean velocity profiles differ only slightly and monotonically approach the
60-second profile. Starting 
with the 90-second profile, the mean
velocities drastically drop. If the time cadence is too short, then granules
have not evolved sufficiently to provide a strong cross-correlation signal. If,
on the other hand, the time cadence becomes too long, then granules have either
evolved too much or completely disappeared so that their horizontal proper
motions are no longer properly measured. Already a time cadence of $\Delta
t=90$~s is a compromise, but many G-band time series are acquired with more than
60~s between successive images. These findings are consistent with a similar
parameter study for \textit{Hinode} G-band images \citep[see Fig. 3
of][]{Verma2011} and independently corroborate that the CO$^\mathrm{5}$BOLD
simulation reproduces essential flow characteristics of solar granulation.

Third, the last important input parameter is the FWHM of the Gaussian sampling
window, which was adjusted in the range from 400 to 1800~km (see
Fig.~\ref{FIG06}), while maintaining a constant time cadence ($\Delta t=60$~s)
and image scale (28~km pixel$^{-1}$). The reference profile
($\mathrm{FWHM}=1200$~km) is once more shown as a black curve. Interestingly,
profiles with small $\mathrm{FWHM}=400$ and 600~km start at high velocities,
then quickly drop, while intersecting all other profiles. These sampling windows
track fine structures with high velocities but the lack of large, coherent
structures (i.e., at least one entire granule) rapidly diminishes the mean flow
speed. The highest flow speeds in the asymptotic part of the mean speed profiles
are found for $\mathrm{FWHM}=800$ and 1000~km. These profiles are very similar
but cut across each other at an averaging time of $\Delta T \approx 14$~min.
Starting at an $\mathrm{FWHM}=1200$~km, the mean speed  profiles are again
systematically arranged without any intersection. Once the sampling window is
sufficiently large to encompass several granules, their stochastic motions tend
to reduce the average flow speed.

In summary, the mean speed profiles in Figs.~\ref{FIG04}, \ref{FIG05}, and
\ref{FIG06} confirm that for an image scale of 80~km pixel$^{-1}$ the LCT input
parameters (averaging time $\Delta T=1$~h, time cadence $\Delta t=60$~s, and a
Gaussian sampling window with an $\mathrm{FWHM}=1200$~km) are indeed the most
reasonable choice to derive horizontal proper motions from \textit{Hinode}
G-band images.

\begin{figure}
\centerline{\includegraphics[width=\columnwidth]{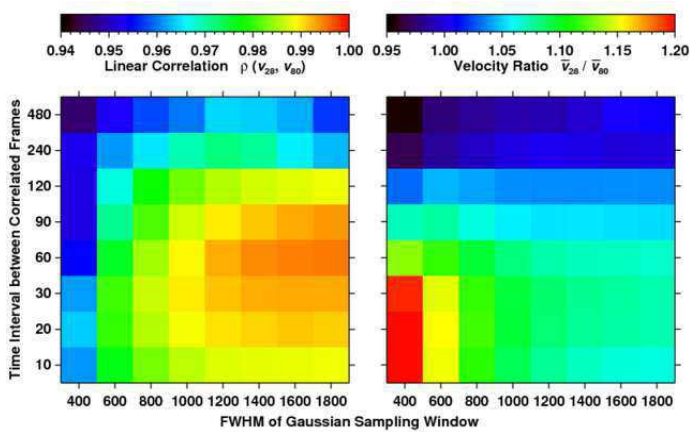}}
\caption{Correlation $\rho ( v_{28}, v_{80})$ (\textit{left}) and mean velocity
ratio $v_{28} / v_{80}$ (\textit{right}) between the flow maps computed from
full-resolution images ($v_{28}$) and from images convolved with a Gaussian of
$\mathrm{FWHM} = 160$~km ($v_{80}$) to match the spatial resolution of
\textit{Hinode} G-band images. For both cases, flow maps were computed with time
cadence $\Delta t = 10$--480~s and Gaussian sampling windows with an
$\mathrm{FWHM} = 400$--1800~km, shown here as $8 \times 8$ square blocks.}
\label{FIG09}
\end{figure}

\begin{figure}
\includegraphics[width=0.48\columnwidth]{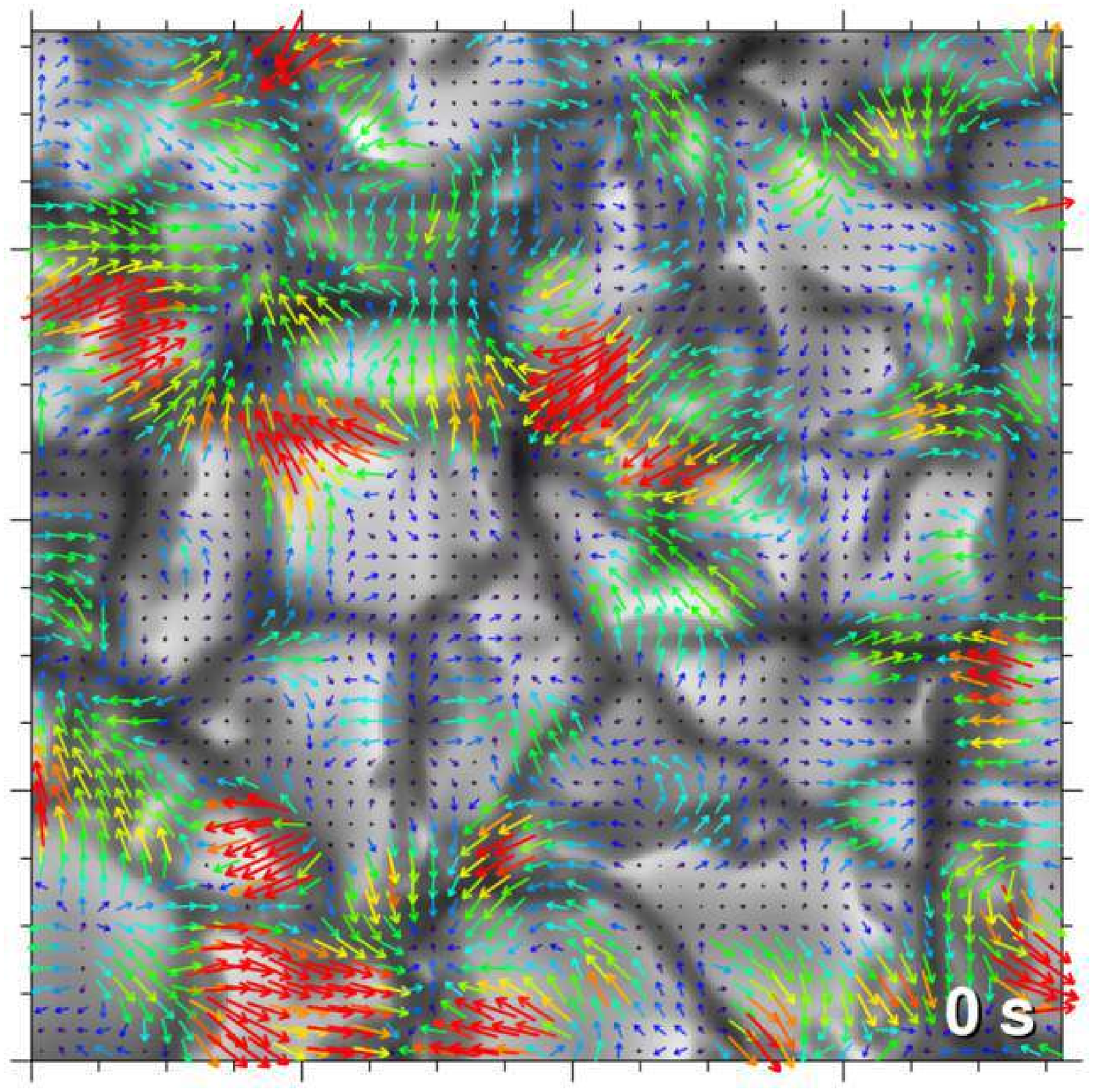}\hfill
\includegraphics[width=0.48\columnwidth]{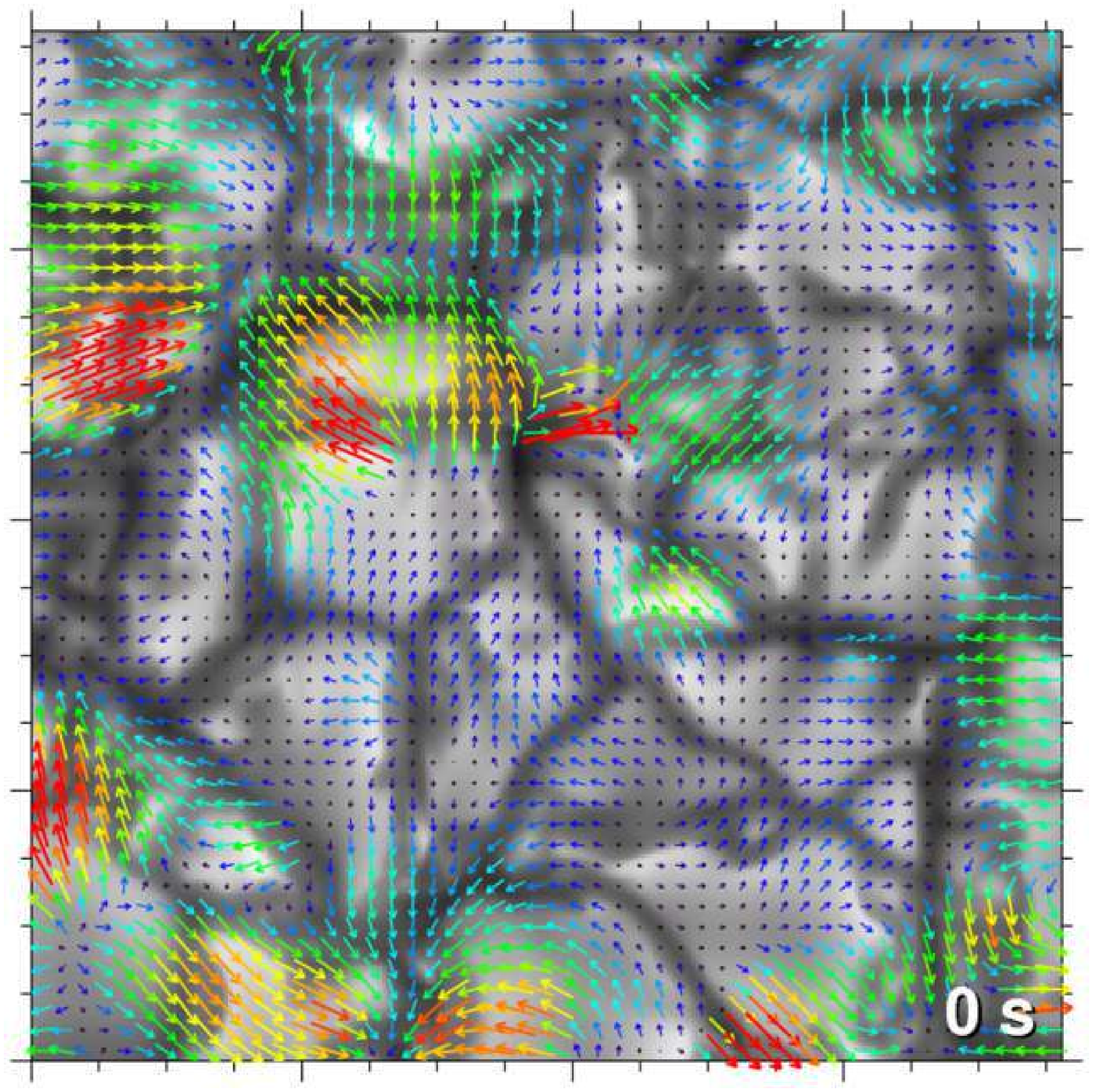}

\smallskip

\includegraphics[width=0.48\columnwidth]{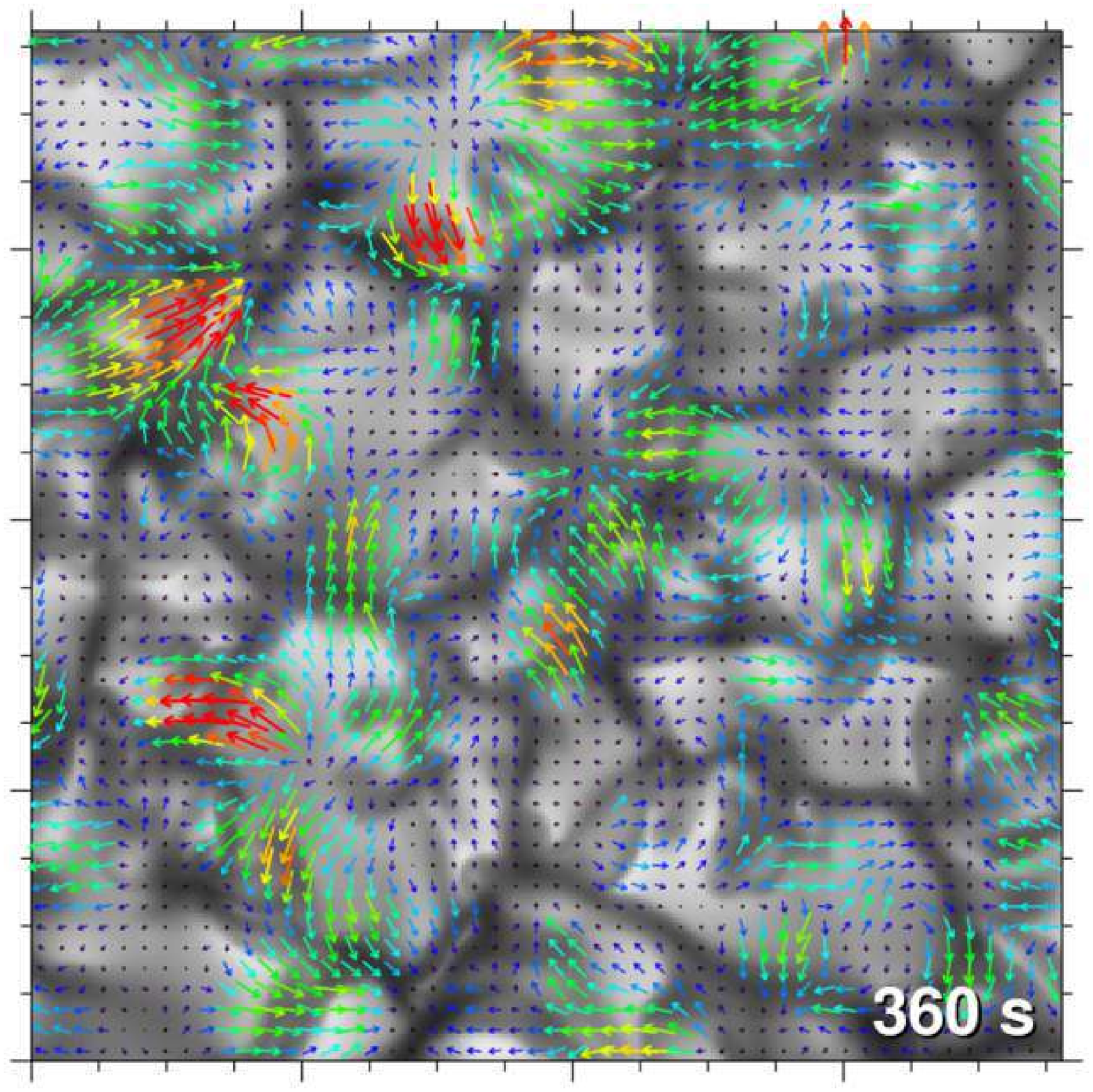}\hfill
\includegraphics[width=0.48\columnwidth]{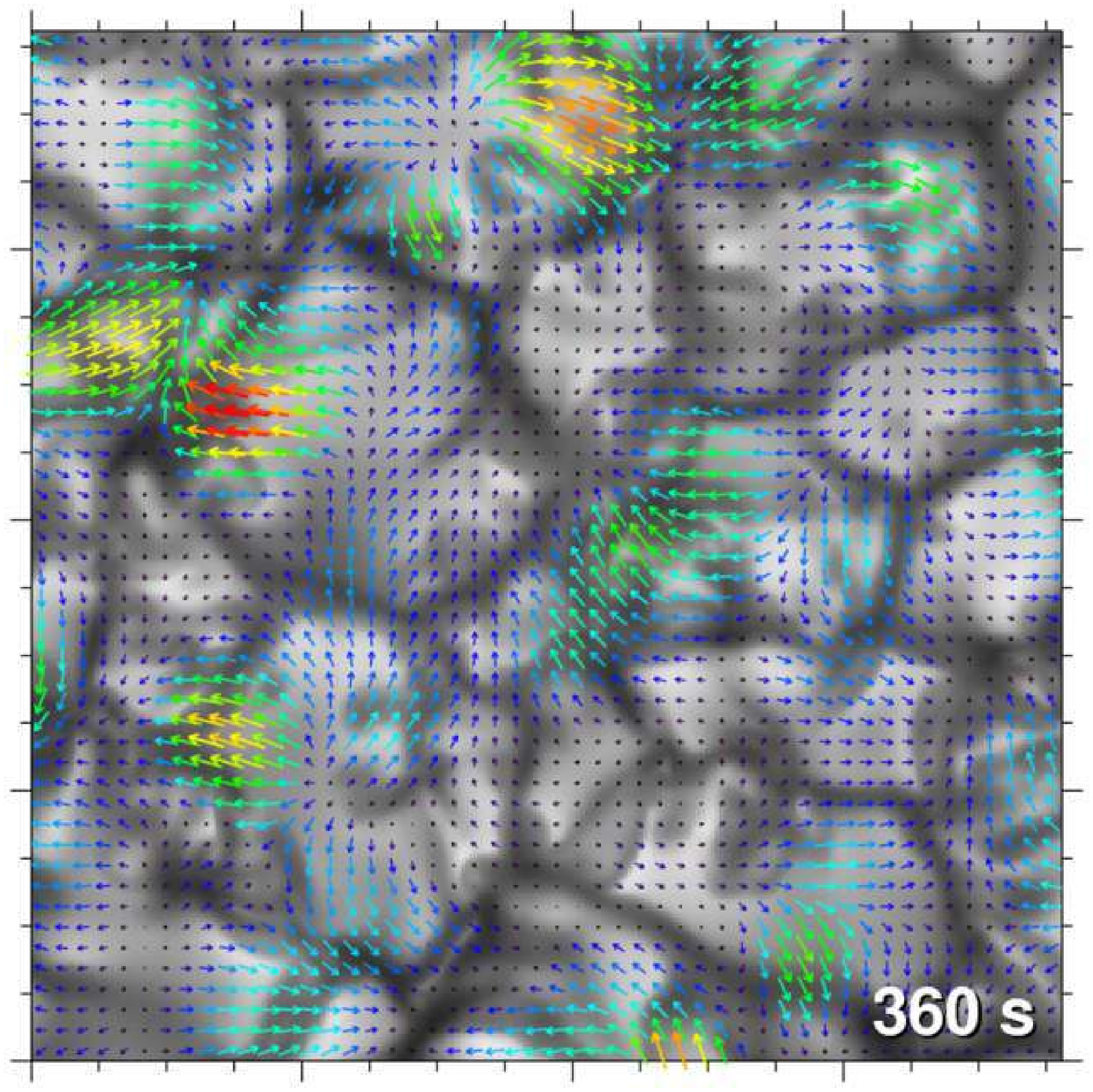}
\caption{Temporal evolution of flow fields for different time intervals between
successive images and sampling window sizes: $\Delta t = 20$~s and
$\mathrm{FWHM} = 600$~km (\textit{left}) and $\Delta t = 60$~s and
$\mathrm{FWHM} = 1200$~km (\textit{right}). The time stamp in the lower right
corner indicates the time elapsed since the beginning of the time series. The
original flow maps were derived from just one pair of correlated images per time
step. These maps with $273 \times 273$ pixels were then resampled to $50 \times
50$ pixels before applying a sliding average over the leading and trailing four
flow maps. Speed and direction of the horizontal proper motions are given by
rainbow-colored arrows (dark blue corresponds to speeds lower than
0.2~km~s$^{-1}$ and red to larger than 2.0~km~s$^{-1}$), which were superposed
on the corresponding gray-scale intensity images of the CO$^\mathrm{5}$BOLD
simulation. Major tickmarks are separated by 2~Mm. The time-dependent LCT flow
fields are shown in Movie~1, which is provided in the electronic edition.}
\label{FIG10}
\end{figure}

\begin{figure*}[t]
\includegraphics[width=\textwidth]{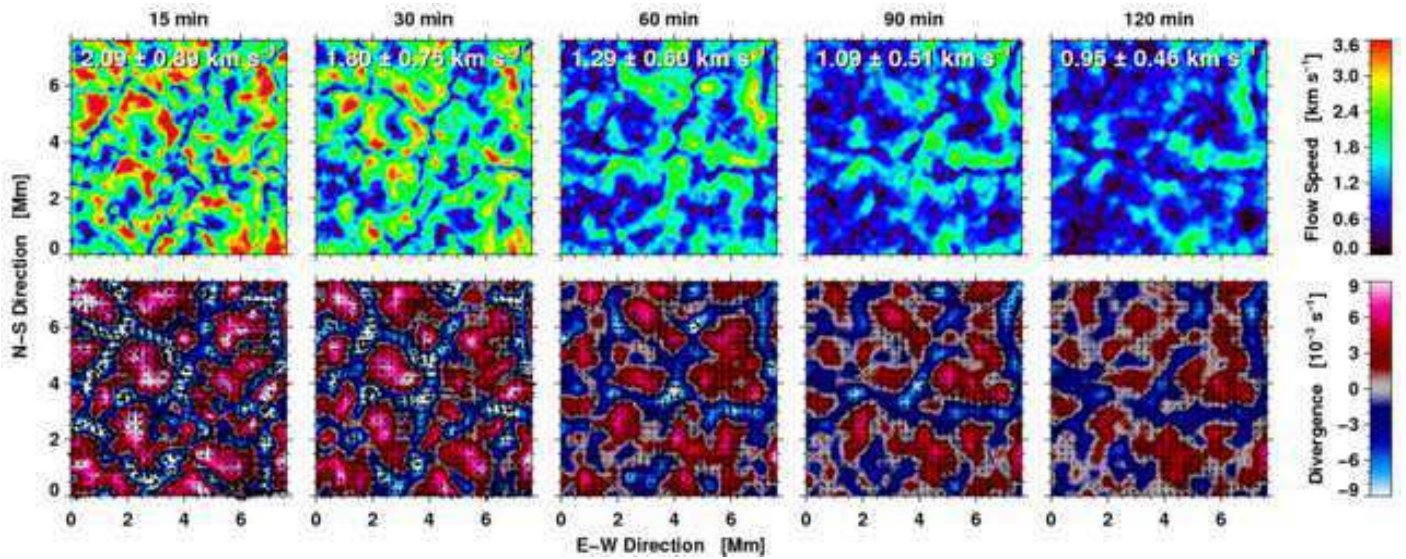}
\caption{Speed (\textit{top}) and divergence (\textit{bottom}) maps averaged
over time intervals of $\Delta T =15$--120~min of the horizontal plasma
velocities corresponding to an optical depth of $\log \tau = 0$. The speed and
divergence values are larger roughly by a factor of three as compared to
Fig.~\ref{FIG03}.}
\label{FIG11}
\end{figure*}

\subsection{Morphology of flow maps and frequency distributions of flow speeds}

Two-dimensional maps of the flow speed offer another approach to evaluate the
influence of the LCT input parameters. In Fig.~\ref{FIG07}, we compiled $8
\times 8$ one-hour averaged speed maps for time cadences of $\Delta t=10$--480~s
and Gaussian sampling windows with an $\mathrm{FWHM}=400$--1800~km. Mean and
standard deviation of the flow speed are given at the top of each panel. This $8
\times 8$ matrix of flow maps facilitates an easy visual comparison of flow
features.

Narrow sampling windows track only small-scale features such as the corrugated
borders of granules, fragmenting granules, and bright points. The flow speeds
are significantly diminished for an $\mathrm{FWHM}=400$~km indicating that
narrow sampling windows miss an important part of the horizontal proper motions
related to granules. If small-scale features are short-lived or travel fast,
significant velocity contributions are only expected for short time-cadences.
Already at a time cadence  $\Delta t=60$~s corresponding velocity signals have
vanished. The same argument related to the lifetime of granules holds for the
longest time cadence that we examined on all spatial scales. At $\Delta
t=480$~s, granules have evolved too much to leave a meaningful cross-correlation
signal. 

Wider sampling windows with short time cadences ($\Delta t=20$--30~s) start to
show higher velocities for an $\mathrm{FWHM}=600$--1200~km. As the FWHM
increases, the flow maps become smoother, and the flow speed starts to decrease.
Broader sampling windows contain more small-scale features exhibiting jumbled
proper motions, and also the number of enclosed granules increases, which
results in weaker flows. Similar results were found for \textit{Hinode} G-band
images \citep[see Fig.~4 and Sect.~4.4 in][]{Verma2011}.
For longer time cadences $\Delta t=120$--240~s, the overall
appearance of the flow maps has not changed but the flow speeds are now much
lower reflecting the lifetime (5--10~min) of solar granulation.

Based on the high resolution images of the CO$^\textrm{5}$BOLD simulations with
an image scale of 28~km pixel$^{-1}$, the ``sweet spot'' with the highest flow
speeds is found for LCT input parameters $\Delta t=20$--30~s and
$\mathrm{FWHM}=800$--1000~km. The parameter pair $\Delta t=60$~s and
$\mathrm{FWHM}=1200$~km yields slightly lower flow speeds but the corresponding
flow maps are virtually identical to the ones within the sweet spot. Considering
the lower spatial resolution (image scale of 80~km pixel$^{-1}$) of
\textit{Hinode} G-band images and the typical cadences of time series of $\Delta
t=60$~s or slightly more, the above parameter pair is still a very good
selection.

The frequency distributions in Fig.~\ref{FIG08} correspond to the $8 \times 8$
flow maps shown in Fig.~\ref{FIG07}. However, the flow speeds were derived from
individual flow maps, i.e., the flow vectors were not averaged before computing
the frequency distributions. This approach might not be suitable for
observational data, which include telescope jitter and seeing.
Thus, only contributions from numerical errors of the LCT method will affect
flow maps based on simulated data.

Apart from the distributions, we calculated median $v_\mathrm{med}$, mean $\bar
v$, $10^\mathrm{th}$ percentile $v_{10}$, and the standard deviation $\sigma_v$
of the speed. The first three values  are depicted in each panel as solid,
long-dashed, and dash-dotted vertical lines, respectively. All the values
describing the frequency distribution are significantly higher than the ones
displayed in Fig.~\ref{FIG07}, because they were obtained from individual flow
maps and not the one-hour averaged data. The first and second moments of the
distributions follow the trend already discussed above for Fig.~\ref{FIG07}. A
high-velocity tail (parametrized by $v_{10}$) and a positive skewness is found
for all distributions, likewise  $\bar v$ is always larger than
$v_\mathrm{med}$.

Each frequency distribution is comprised of more than 25 million flow vectors
and no smoothing was applied. Thus, not only their overall shape but also the
minute detail is significant. Small ripples become visible around the maximum of
the distributions for $\Delta t=480$~s and an $\mathrm{FWHM} \ge 1200$~km,
suggesting that the initial feature has evolved and is no longer tracked, or
other features are tracked instead. Furthermore at $\Delta t=480$~s, the
high-velocity tail has vanished in all distributions, most prominently for
shorter time cadences. Another interesting feature is the shoulder at the
high-velocity side of the distributions for short time cadences and narrow
sampling windows, which hints at a contribution from smaller features with
higher velocities. Exploding or fragmenting granules and bright points, thus,
might have a different velocity spectrum distinguishing them from regular
granules. In summary, Figs.~\ref{FIG07} and \ref{FIG08} can also be taken as a
point of reference for many other LCT studies with various input parameters,
hopefully providing a more cohesive description of horizontal proper motions in
the photosphere and chromosphere, or where ever time-series of images are
available.

As a corollary to the above parameter study, we repeated the LCT measurements
but now with smoothed simulation data  (Gaussian kernel with an $\mathrm{FWHM}
=160$~km) to have the same spatial resolution (image
scale of 80~km pixel$^{-1}$) as \textit{Hinode} G-band images. In
Fig.~\ref{FIG09}, we use the linear correlation coefficient  $\rho$($v_{28}$,
$v_{80}$) and the ratio of the velocities $v_{28} / v_{80}$ for both
image scales to quantify how a coarser spatial resolution
affects the flow maps. The correlation coefficients are lowest for narrow
sampling windows ($\mathrm{FWHM}=400$~km) and long time cadences $\Delta t \ge
240$~s because much of the fine structure (with high velocities) has been
blurred. The highest correlations are found for $\mathrm{FWHM} \ge 1200$~km and
$\Delta t= 20$--90~s. Only a very small deviation of less than 0.001 from a
perfect correlation is observed for the LCT input parameters chosen in
\citet{Verma2011}. 

The right panel in  Fig.~\ref{FIG09} demonstrates that flow speeds could be
underestimated by as much as 20\% in the case of G-band images (neglecting so far
the comparison with the plasma velocities in Sect.~4.5). However, such strong
deviations are only observed for very narrow sampling windows and very short
time cadences. In general, the speeds in both cases differ by less than 5\%.
Interestingly, if the time cadence is long, then there are not many traceable
features left in the sampling window, but by additional smoothing, more coherent
features are created that are long-lived, which explains the slightly higher
velocities $v_{80}$ for G-band images in the $\Delta t = 240$--400~s range.
Considering that HMI continuum images have an image scale of about 360~km
pixel$^{-1}$ and a cadence of  $\Delta t = 45$~s, broad sampling windows
($\mathrm{FWHM} \ge 1600$~km) might be needed to build up a reliable
cross-correlation signal. Even though the flow speeds might be underestimated by
10--20\% as compared to high-resolution simulation data (see Figs.~\ref{FIG07}
and \ref{FIG08}), the overall morphology of the flow field will still be
reliably recovered with correlation coefficients close to unity (see left panel
of  Fig.~\ref{FIG09}). This was also demonstrated by \citet{Beauregard2012} for
complex flows along the magnetic neutral line of active region NOAA~11158 at the
time of an X2.2 flare.

\begin{figure}
\centerline{\includegraphics[width=\columnwidth]{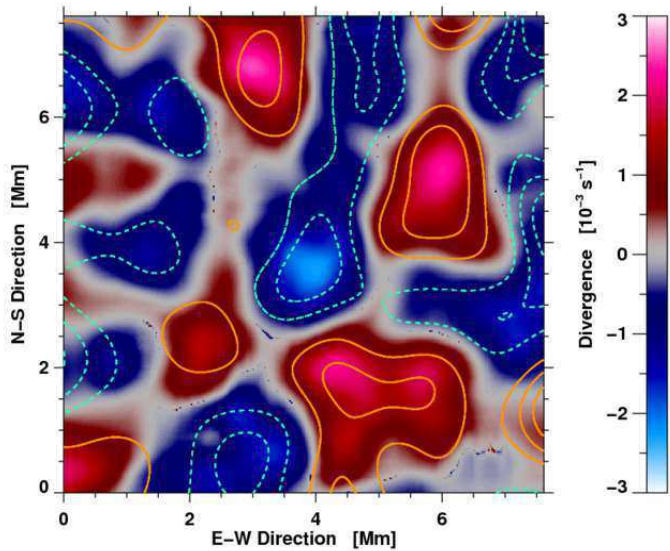}}
\caption{LCT divergence map computed for input parameters: image scale 80~km
pixel$^{-1}$, time cadence $\Delta t=60$~s, Gaussian sampling window with an
$\mathrm{FWHM}=1200$~km, and averaging time $\Delta T=1$~h. Overplotted are the
contours of the corresponding one-hour averaged divergence of the actual flow
velocity smoothed by a Gaussian with an $\mathrm{FWHM}=1266$~km. Orange
(\textit{solid}) and blue (\textit{dashed}) lines indicate positive and
negative ($\pm 1$, $\pm 2\, \mathrm{and}\ \pm 3 \times10^{-3}$~s$^{-1}$)
divergence, respectively.}
\label{FIG12}
\end{figure}

\subsection{Dynamics of horizontal proper motions}

Besides quantifying persistent flow fields, LCT techniques can also capture the
dynamics of horizontal proper motions. In Fig.~\ref{FIG10}, we compare snapshots
from two movies with two different sets of LCT input parameters: $\Delta t =
20$~s and  $\mathrm{FWHM} = 600$~km (left column) and $\Delta t = 60$~s and
$\mathrm{FWHM} = 1200$~km (right column). The snapshots are separated by 360~s in
time and show the continuum intensity with a $50 \times 50$ grid of superposed,
color-coded flow vectors. The flow maps were smoothed both in space ($5 \times
5$-pixel neighborhood) and time (sliding average of nine individual flow maps).
These values were chosen such that watching the movies leaves a smooth and
continuous visual impression. Less smoothing will result in jittering arrows in
some places. We counterpoint the narrow sampling window/high cadence case with
our typical choice of LCT input parameters. In the first case, much higher flow
speeds are apparent, and the flow vectors in regions with high flow speeds show
indications of vortex or twisting motions, which are absent in the latter case,
where the flow vectors are more uniformly arranged. Comparing the flow fields at
0~s and 360~s clearly demonstrates that periods with many strong flow kernels
are followed by much quieter flow fields. The high flow speeds have their origin
in the borders of rapidly expanding or fragmenting granules.

\begin{figure}
\includegraphics[width=\columnwidth]{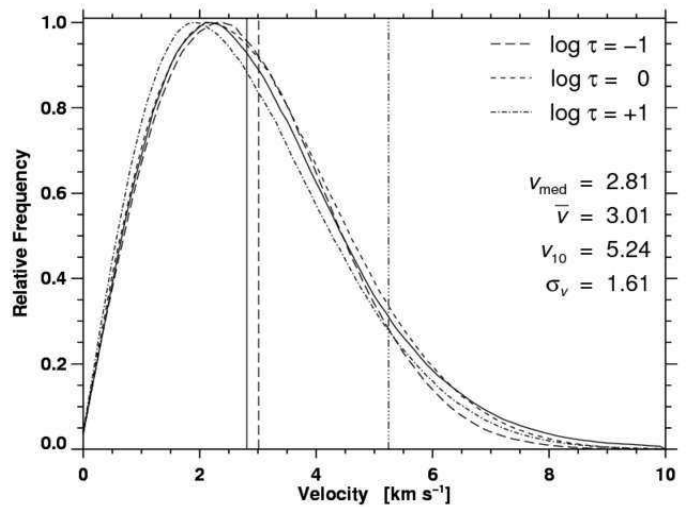}
\caption{Relative frequency distributions for the horizontal plasma velocities
corresponding to different optical depths of $\log \tau= -1$, 0, and $+1$, which
are depicted as long-dashed, dashed, and dash-dotted curves, respectively. The
three vertical lines mark the position of median $v_\mathrm{med}$
(\textit{solid}), mean $\bar {v}$ (\textit{long-dashed}), and $10^\mathrm{th}$
percentile $v_{10}$ (\textit{dash-dotted}) values of speed at an optical depth
of $\log \tau= 0$. A frequency distribution (\textit{solid}) for LCT flow
velocities was stretched by a factor of about three so that the root-mean-square
deviation from the profile with $\log \tau = 0$ was minimal. The LCT input
parameters were image scale 80~km pixel$^{-1}$, time cadence $\Delta t = 60$~s,
and a Gaussian sampling window with an $\mathrm{FWHM} =1200$~km.}
\label{FIG13}
\end{figure}

\subsection{Comparison with plasma velocities}

The actual plasma velocities are given for three optical depths $\log \tau =
-1$, 0, and $+1$. The time dependence of the average flow field along with the
corresponding divergence maps are depicted in Fig.~\ref{FIG11} for plasma
velocities at a depth of $\log\tau =0$. Both speed and divergence values are
higher by a factor of two to three as compared to the maps in Fig.~\ref{FIG03}.
The highest factor between average LCT and plasma velocities is encountered for
shorter time averages. For averaging times of $\Delta T = 15$--30~min, the speed
maps exhibit more fast-moving, small-scale features associated with the
boundaries of expanding or fragmenting granules. Increasing the averaging time
results in smoother speed maps with much reduced speed values. By visually
comparing Figs.~\ref{FIG03} and \ref{FIG11}, it becomes apparent that
significant spatial smoothing has to be applied to the speed and divergence maps
of the plasma velocities. This is a direct consequence of the sampling windows
employed in LCT.

\begin{figure}
\includegraphics[width=\columnwidth]{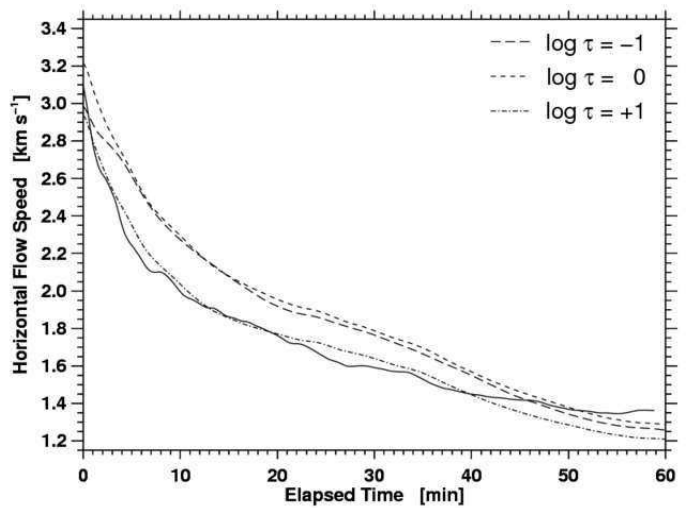}
\caption{Average horizontal flow speeds as a function of the elapsed time for
the horizontal plasma velocities corresponding to optical depths of $\log \tau =
-1$, 0, and $+1$ are depicted as long-dashed, dashed, and dash-dotted curves,
respectively. The solid curve, as in Fig.~\ref{FIG13}, has to be multiplied by a
factor of three to match the profile for $\log\tau = +1$.}
\label{FIG14}
\end{figure}

Figure~\ref{FIG12} shows the divergence map for LCT flow fields computed for a
one-hour time-series of images, which were smoothed to match the spatial
resolution of \textit{Hinode} G-band images, a time cadence of $\Delta t=60$~s,
and a Gaussian sampling window with an $\mathrm{FWHM}=1200$~km. We overplotted
the divergence of the actual flow velocity from an optical depth of $\log \tau =
0$ averaged over one hour and smoothed it with a Gaussian kernel with an
$\mathrm{FWHM}=1266$~km. The size of the smoothing kernel was chosen such that
linear correlation between the LCT and plasma maps was maximized ($\rho =$0.90).
The position of positive and negative divergence corresponding to the plasma
flows roughly matches with the LCT divergence extrema. Even though the
correlation between LCT and plasma divergence maps is significant (once properly
smoothed), there are still morphological differences, not to mention the drastic
difference in the absolute values of speed and divergence. In general, our
findings are in good agreement with \citet{Matloch2010}, e.g., their Fig.~2. In
Fig.~\ref{FIG09}, we found a negligible dependence of the LCT results on the
image scale (28 vs.\ 80~km pixel$^{-1}$) or in this context equivalently the
spatial resolution. Hence, for the further discussion, we use G-band-type LCT
flow maps.

\begin{figure}
\centerline{\includegraphics[width=\columnwidth]{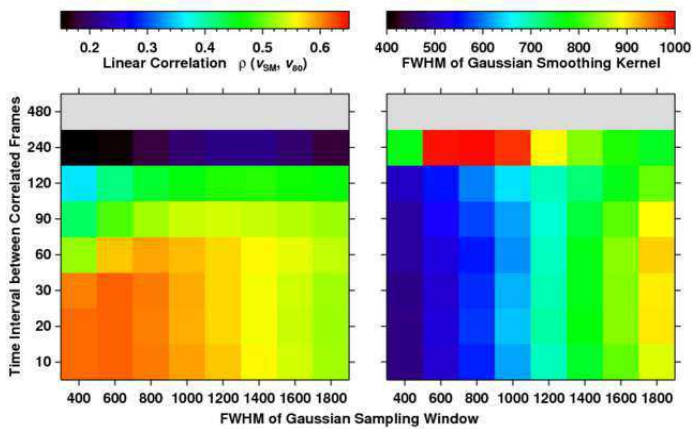}}
\caption{Maximum correlation $\rho(v_\mathrm{sm},v_{80})$ (\textit{left}) and
corresponding FWHM of Gaussian kernel (\textit{right}) used to smooth the
actual flow map $v_\mathrm{sm}$. For both cases, the maps were averaged over one
hour. The LCT flow maps were computed for images with an image-scale of 80~km
pixel$^{-1}$ matching \textit{Hinode} G-band images, using time cadences of
$\Delta t=10$--480~s and Gaussian sampling windows with an
$\mathrm{FWHM}=400$--1800~km, shown here as $8 \times 8$ square blocks. Gray
blocks indicate parameters, where the linear correlation did not deliver
meaningful results.}
\label{FIG15}
\end{figure}

The frequency distributions of the actual plasma velocities are shown in
Fig.~\ref{FIG13}. The distributions are assembled from the 180 velocity maps
covering about one-hour. The mode of the distributions is shifted to higher
velocities for higher atmospheric layers. For comparison, we show the LCT-based
velocity distribution. The LCT input parameters were a time cadence of $\Delta t
= 60$~s and a Gaussian sampling window with an $\mathrm{FWHM} = 1200$~km.
Similar to Fig.~\ref{FIG08}, the flow speeds were derived from the individual
flow maps, i.e., the flow vectors were not averaged before computing the
frequency distributions. We scaled the LCT frequency distribution by the factor
of $\approx 3.01$ in velocity to match it with the distributions for the actual
plasma velocities. The stretched LCT and actual velocity distributions have very
similar shapes. For an optical depth of $\log \tau = 0$, it had the lowest
${\chi}^2$-error. However, the differences in the ${\chi}^2$-error for all three
optical depths are not significant.

In Figure~\ref{FIG14}, we plotted the mean flow speed as a function of elapsed
time for the actual plasma velocities at optical depths $\tau = -1$, $0$, and
$+1$. We overplotted the same relation for the  LCT-computed flow speed. This
curve was scaled by a factor of $\approx 2.85$ to match it with the curve for
optical depth $\log \tau = +1$ as it had the lowest ${\chi}^2$-error. The
profiles for plasma and scaled LCT velocities are very similar up to about
40~min, then the LCT velocities level out at higher velocities, which can be
attributed to accumulating numerical errors of the LCT algorithm but also to
small drifts over time of the intensity images. However, the fact that the
curves track each other for the first 40~min is a strong indicator that LCT
reliably detects the time-dependence of the actual plasma flows. Although the
differences in the ${\chi}^2$-error are of the same order of magnitude for all
three optical depths, the lower ${\chi}^2$-error at an optical depth of $\log
\tau = +1$ indicates that LCT is picking up velocities at deeper layers in the
photosphere.
 
\citet{Matloch2010} used the same Gaussian kernel for smoothing the plasma flows
as they used for computing the LCT flow maps. In principle, this is a free
parameter, which can be varied to determine how much smoothing is required to
get the highest correlations between plasma and LCT flow maps. We smoothed the
one-hour averaged actual flow speed $v_\mathrm{sm}$ and divergence maps  with a
Gaussian kernel of size 64 $\times$ 64 pixels and with varying
$\mathrm{FWHM}=100$--1500~km. We correlated it with the one-hour averaged LCT
speed $v_{80}$ and divergence maps for time cadences $\Delta t=10$--480~s and
Gaussian sampling windows with an $\mathrm{FWHM}=400$--1800~km. We performed
this smoothing-correlation evaluation for plasma speeds at all three optical
depths. However, only the results for the speed corresponding to optical depth
$\log \tau=+1$ and the corresponding FWHM of the Gaussian smoothing kernel are
shown in Fig.~\ref{FIG15}. The top rows for both correlation coefficient and
FWHM are empty because we found no meaningful correlation between LCT maps
computed with a time cadence of $\Delta t=480$~s and the smoothed plasma flow
maps. One can notice this lack of correlation already in the row for the time
cadence $\Delta t = 240$~s, which shows a different behavior as compared to the
remaining rows. This strongly substantiates earlier results indicating that time
cadences in excess of 4~min are too long to deliver meaningful LCT results in
the case of photospheric continuum images.

\begin{figure}
\begin{center}
\includegraphics[width=0.45\columnwidth]{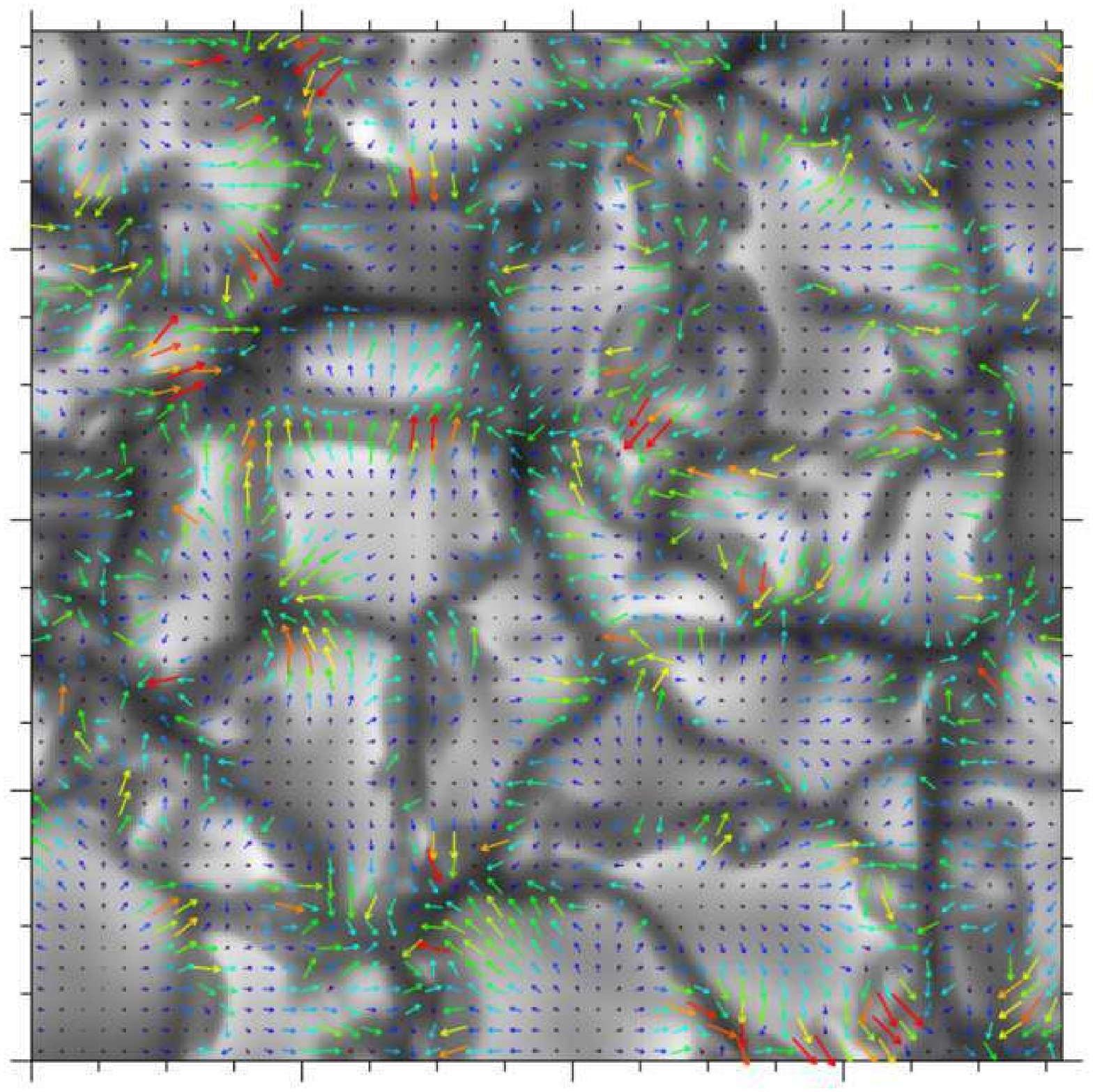}
\hspace*{2mm}
\includegraphics[width=0.45\columnwidth]{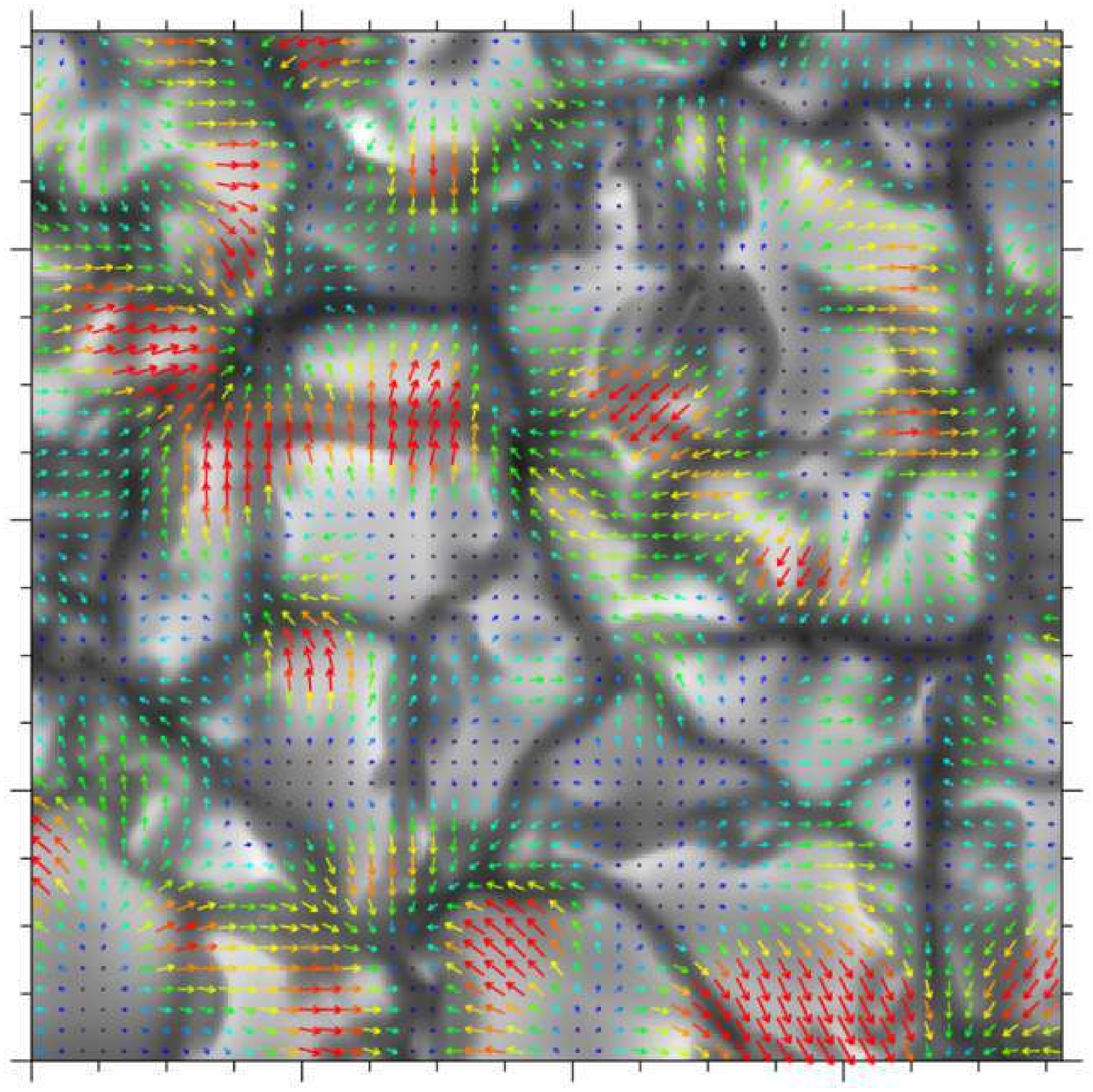}
\end{center}
\caption{First gray-scale intensity image of the time series overplotted with
the actual (\textit{left}) and smoothed CO$^5$BOLD plasma velocity vectors at
$\log\tau = +1$ (see left most panel in Fig.~\ref{FIG10} for the corresponding
LCT flow vectors). The flow field was smoothed using a Gaussian kernel with an
$\mathrm{FWHM}=686$~km (see Fig.~\ref{FIG15}) corresponding to the maximum
correlation coefficient between plasma and LCT (time cadence $\Delta t = 60$~s
and Gaussian sampling window with an $\mathrm{FWHM} = 1200$~km) velocities. The
speed and direction of the flow field are given by rainbow-colored arrows, where
dark blue corresponds to low and red to high velocities within the range of
0.75--7.5~km~s$^{-1}$ and 0.3--3.0~km~s$^{-1}$ for the actual and smoothed flow
fields, respectively. The time-dependent CO$^5$BOLD flow fields are shown in
Movie~2, which is provided in the electronic edition.}
\label{FIG16}
\end{figure}

The mean correlations $\bar \rho$ between the smoothed plasma $v_\mathrm{sm}$
and LCT $v_{80}$ velocities are 0.25, 0.41, and 0.49 at optical depths of
$\log\tau = -1$, 0, and $+1$, respectively. The highest correlations are found
at $\log\tau = +1$ indicating that LCT speed correlates better with plasma flows
emerging from deeper layer irrespective of the LCT input parameters. Speed maps
show the best correlations for smaller time cadences and narrow sampling
windows. The actual plasma velocities were smoothed with kernels that are
in general narrower than the LCT sampling windows (the only exception arises for
the smallest $\mathrm{FWHM} = 400$~km). 

The divergence maps again match best with flows from deeper layers. However, the
FWHM of the Gaussian kernels are about twice the size of kernels used for speed
maps. At the moment, we do not have a physical interpretation why much stronger
smoothing has to be applied to the plasma divergence.

All our previous findings and the difficulties in interpreting LCT flow and
divergence maps have their origin in the instantaneous flow field
\citep[as already pointed out by][]{Rieutord2001}. In the left
panel of Fig.~\ref{FIG16}, the local CO$^5$BOLD plasma velocities ($\log\tau =
+1$) are superposed as color-coded arrows on top of the first continuum image in
the time series. Only minor spatial smoothing over a $5 \times 5$-pixel
neighborhood was applied. The strongest horizontal flows are encountered near
the boundaries of the expanding and fragmenting granules. The center of granules
is almost always a divergence center and the velocity at these locations is
significantly reduced. The visual appearance of this flow map is quite different
from what is found in the LCT analysis. In the right panel  of Fig.~\ref{FIG16},
we smoothed the CO$^5$BOLD plasma velocities using a Gaussian kernel with an
$\mathrm{FWHM} = 686$~km, which produced the highest linear correlation $\rho
\approx 0.57$ for the time cadence $\Delta t = 60$~s and a Gaussian sampling
window with an $\mathrm{FWHM} = 1200$~km. As a result, velocity vectors
reflecting different plasma properties (high/low speeds and convergent/divergent
motions) become intermingled. The final flow map is additionally modulated by
the morphology of the granular cells, which are included in the sampling window.
Even by using narrower sampling windows and higher time cadences it is doubtful
that the original plasma velocities can be recovered, once having been scrambled
in the LCT algorithm. On the other hand, any average property of the plasma
flows, which does not depend on the spatial location should remain unaffected
(see Figs.~\ref{FIG13} and \ref{FIG14}).


\section{Conclusions}

Applying LCT techniques to simulated data of solar granulation has proven as an
excellent diagnostic to evaluate the performance of the algorithm. By
contrasting LCT velocities with the plasma velocities of the simulation, some of
the inherent troubles with optical flow techniques became apparent. Based on the
previous analysis, we draw the following conclusions:
(1) One-hour averaged LCT flow and divergence maps differ
significantly, if separated by more than one hour in time. As expected, the
simulation of
granulation does not contain any systematic persistent flow features. However,
some strong flow kernels might still survive the averaging process but they can
still be of intermittent nature.
(2) The time over which individual flow maps are averaged
critically determines if the LCT flow field reflects either instantaneous
proper motions by individual granules or longer lasting flow features. The
functional dependence of the mean flow speed on the elapsed time indicates that
an averaging time of at least 20~min (several times the life time of a granule)
is needed to raise persistent flows to the state of being significant.
(3) Time cadences $\Delta t=240$--480~s are not suitable for
tracking photospheric continuum images because the features (granules) in the
tracking window have either evolved too much or moved outside the sampling
window so that the cross-correlations become meaningless. (4) Smaller sampling 
windows track fast-moving fine structures, if
high-cadence images are available, but lack the ability to measure horizontal
proper motions of coherent features, which results in an underestimation of the
flow speed. However, the morphology of the flow field is recovered better with
narrower sampling windows.
(5) Frequency distributions of flow speeds for short time cadences
and narrow sampling windows indicate that exploding or fragmenting granules and
bright points have a different velocity spectrum distinguishing them from
regular granules.
(6) LCT yields the highest speed values for sampling windows with an
$\mathrm{FWHM}=800$--1000~km and short time cadences $\Delta t=20$--30~s. (7)
The input parameters time cadence $\Delta t=60$~s, Gaussian sampling window with
a $\mathrm{FWHM}=1200$~km, and averaging time $\Delta T=1$~hr are the most
reasonable choice for bulk-processing of \textit{Hinode} G-band images.
(8) Both the stochastic nature of granulation and the choice of
LCT input parameters might be responsible for the often conflicting values in
literature concerning flow speed, divergence, and vorticity.
(9) Significant smoothing has to be applied to the actual plasma
velocities to match the LCT flow and divergence maps. The typical velocity
pattern -- high velocities at the border and low speeds in the center of
granules -- will however be lost in the smoothing process. Thus, even with very
narrow sampling windows, short time cadences, and images free of aberrations and
distortions, recovering details of the plasma flows might proof a futile
undertaking.
(10) The LCT speeds are underestimated by a factor of three as
is evident from the LCT frequency distributions (Fig.~\ref{FIG13}) and the
curve for the mean LCT flow speed (Fig.~\ref{FIG14}). This scaling factor
does not mean that LCT speeds can simply be multiplied by
it, rather it serves as a reminder that flow fields derived from LCT
have to be interpreted with caution: spatial smoothing has to be carefully
considered and deriving flow velocities from intensity images will not always
reflect the underlying plasma velocities.

Although the current study focuses only on granulation, in principle a similar
study could be performed based on simulated data of active regions. Realistic
MHD simulations of sunspots \citep[e.g.,][]{Rempel2009, Cheung2010} could
provide the basis for such a study, which will potentially aid in the
interpretation of persistent flow features such as the divergence line in the
mid-penumbra \citep[e.g.,][]{Denker1998b} or the distinct flow channels of
moving magnetic features connecting the sunspot's penumbra to the surrounding
supergranular cell boundary \citep{Verma2012a}.


\begin{acknowledgements}
M.V. expresses her gratitude for the generous financial support by the German
Academic Exchange Service (DAAD) in the form of a Ph.D. scholarship. C.D. was
supported by grant DE 787/3-1 of the German Science Foundation (DFG).
\end{acknowledgements}


\bibliographystyle{../../LaTeX/aa}
\bibliography{../../LaTeX/an-jour,../../LaTeX/meetu}

\end{document}